# A Novel Data Hiding Scheme for Binary Images

Do Van Tuan
Hanoi College of Commerce and Tourism
Hanoi – Vietnam
dvtuanest@gmail.com

Tran Dang Hien
Vietnam National University
hientd_68@yahoo.com

Pham Van At
Hanoi University of Communications and Transport
phamvanat83@vnn.vn

*Abstract* - **this paper presents a new scheme for hiding a secret message in binary images. Given m×n cover image block, the new scheme can conceal as many as $\lfloor log_2(m \times n + 1) \rfloor$ bits of data in block, by changing at most one bit in the block. The hiding ability of the new scheme is the same as Chang et al.'s scheme and higher than Tseng et al.'s scheme. Additionally, the security of the new scheme is higher than the two above schemes.**

## I. INTRODUCTION

Nowadays, the Internet is the most popular channel for data exchanges between providers and users. Yet, the data safety issue on the Internet is always a challenge to managers and researchers, as the data on the Internet is easily tampered with and stolen by hackers during transmission. In addition to encryption schemes, data hiding has an important role in secret message transmission, authentication, and copyright protection on public exchange environment.

Data hiding is a technique to conceal a secret message in cover media, to avoid arousing an attacker's attention. The cover media is often a document, image, audio or video. According to [1], the data hiding schemes proposed in an image can be divided into two categories. In the first category, the schemes hide a secret message in the spatial domain of the cover image [3,4,6,] and the least significant bits of each pixel in cover image is modified to hide the secret message. In the second category, the schemes hide a secret message in transformed domain of cover image [2,8]. Several transformation functions, such as discrete cosine transform and discrete wavelet transform are widely used.

However, most cover images of the above schemes are gray-level images or color images. The binary image is not often used in cover media [1,5,7]. The major reason is that the modification is easily detected when a single pixel is modified in a binary image. For binary images, two schemes are seen as modern and efficient in TCP scheme [5] and CTL scheme [1]. Accordingly, given an m×n cover image block from cover image, both schemes can conceal maximum $r = \lfloor log_2(m \times n + 1) \rfloor$ bits in block. To hide r bits, TCP scheme changes two pixels at most, but CTL scheme only need change one pixel at most. Therefore, the invisibility of CTL scheme is higher than TCP scheme. However, the content of the CTL scheme is quite complicated. This paper presents a novel scheme to hide a secret message in binary images. In addition, the hiding capacity and stego-image quality of new scheme are the same with CTL scheme, but the security property of the new scheme is higher than CTL scheme. Moreover, the content of new scheme is simpler than above two schemes.

The remaining text of this paper is organized as follows: In section 2, we define some operators used in this paper. In section 3, we present some hiding data algorithms in a block. These algorithms are background for new data hiding scheme presented in section 4. In section 5, we present some experimental results. Finally, Section 6 presents the conclusions.

## II. NOTATION

**Definition 1.** Denote $\otimes$ is component-wise multiplication of two matrices of the size m×n:

$$(F \otimes G)_{i,j} = F_{i,j} \times G_{i,j} \text{ , } i = 1,2,\ldots,m \text{ and } j = 1,2,\ldots,n$$

**Definition 2.** Denote $\oplus$ is bit-wise XOR operator on two nonnegative integers

Example: $5 \oplus 12 = 0101 \oplus 1100 = 1001 = 9$

**Definition 3.** For every nonnegative integer matrix D, XSUM(D) or $\sum_{i,j}^{\oplus} D_{i,j}$ is the sum by operator $\oplus$ over all component of D.

**Remark 1**. If $D_{i,j} \in \{0,1,\ldots,2^r - 1\} \forall (i,j)$, then

$$XSUM(D) \in \{0,1,\ldots,2^r - 1\}$$

## III. HIDING DATA ON ONE BLOCK

This section presents algorithms for hiding data on a binary matrix (block of pixels) F of size m×n by modifying one bit at most in F.

### A. Algorithm for hiding one bit

Wu-Lee scheme [7] is known as a simple scheme for hiding data on binary images. This scheme uses a binary random matrix K of size m×n as secret key and can hide a bit b on F by modifying one bit at most of F to get a binary matrix G to satisfy the condition:

$$SUM(G \otimes K) \mod 2 = b$$

However, this scheme can not extend to hide a string of bits. Now, we consider a new algorithm by using operator $XSUM(G \otimes K)$ instead of $SUM(G \otimes K)$ in the Wu-Lee algorithm. This algorithm could expand to hide a string of r bits.

**Algorithm 1.**

This algorithm will modify at most one element of F to get a matrix G satisfying the condition:

$$XSUM(G \otimes K) = b$$

Algorithm is performed as follows:

**Step 1:**
- Compute $s = XSUM(F \otimes K)$
- If s=b then set G=F and stop

    Otherwise go to Step 2

**Step 2:**
- Compute $d = s \oplus b$
- Find an element (u,v) such that $K_{u,v} = d$
- Reverse $F_{u,v}$: $F_{u,v} = 1 - F_{u,v}$
- Set G = F and stop

**Remark 2.** The value of d is always equal to 1, so to Step 2 are carried out, the matrix K must satisfy the condition:

$$\{1\} \subset \{K_{i,j} | i = 1, \ldots, m \text{ and } j = 1, \ldots, n\}$$

*B. Algorithm for hiding a bit string*

In this section we expand the Algorithm 1 for hiding r bits $b = b_1 b_2 \ldots b_r$ in an image block F by using the matrix P for which elements are strings of r bits. In other words, the elements $P_{i,j}$ have a value from 0 to $2^r - 1$.

Similar to the Algorithm 1, following algorithm will change at most one element of the matrix F to obtain matrix G to satisfy the condition:

$$XSUM(G \otimes P) = b \quad (3.1)$$

**Algorithm 2.**

**Step 1:**
- Compute $s = XSUM(F \otimes P)$ (3.2)
- If s = b, set G = F and stop

    Otherwise go to Step 2

**Step 2:**
- Compute $d = s \oplus b$
- Find an element (u,v) such that $P_{u,v} = d$
- Reverse $F_{u,v}$: $F_{u,v} = 1 - F_{u,v}$
- Set G = F and stop

**Remark 3.** In the above algorithm, the value of d is an integer number from 1 to $2^r - 1$, so to Step 2 are carried out, the matrix P must satisfy the condition:

$$\{1, \ldots, 2^r - 1\} \subseteq \{P_{i,j} | i = 1, \ldots, m \text{ and } j = 1, \ldots, n\} \quad (3.3)$$

From the condition (3.3) it follows that

$$r \leq \lfloor \log_2(m \times n + 1) \rfloor$$

*C. Example*

To illustrate the contents of Algorithm 2, we consider an example for which $b = b_1 b_2$ and matrices F, P are defined as follows:

**b=$b_1 b_2$ =10**

F

| 1 | 0 | 0 |
|---|---|---|
| 0 | 1 | 1 |
| 0 | 1 | 1 |

P

| 10 | 01 | 00 |
|----|----|----|
| 11 | 01 | 10 |
| 11 | 11 | 01 |

Step 1:
- $s = XSUM(F \otimes P) = 10 \oplus 01 \oplus 10 \oplus 11 \oplus 01 = 11$
- Since s ≠ b, go to Step 2.

Step 2:
- $d = s \oplus b = 11 \oplus 10 = 01$
- Find (u,v) for which $P_{u,v} = d = 01$. In this case, we have three choices: (1,2), (2,2) and (2,3). Choose (u,v)=(1,2)
- Reverse $F_{1,2}$: $F_{1,2}$=1-0 = 1, and set G = F.

So after hiding two bits 10 on F, we obtain G as follows:

G

| 1 | 1 | 0 |
|---|---|---|
| 0 | 1 | 1 |
| 0 | 1 | 1 |

*D. Correctness of the data hiding scheme*

We need to prove matrix G obtained from Algorithm 2 satisfies condition (3.1): $XSUM(G \otimes P) = b$. This is obviously true if the algorithm ends in Step 1, so we only consider the case of the algorithm ends at step 2. Then we have:

$$P_{u,v} = d = s \oplus b \in \{1, 2, \ldots, 2^r - 1\} \quad (3.4)$$

$$G_{i,j} = \begin{cases} F_{i,j}, & \text{if } (i,j) \neq (u,v) \\ 1 - F_{i,j}, & \text{if } (i,j) = (u,v) \end{cases} \quad (3.5)$$

Now if set

$$s' = XSUM(G \otimes P) = \sum_{i,j}^{\oplus} G_{i,j} \times P_{i,j}$$

Then from (3.2), (3.5) and from the fact that $a \oplus a = 0$, we obtain

$$s' = \sum_{(i,j) \neq (u,v)}^{\oplus} F_{i,j} \times P_{i,j} \oplus [(1 - F_{u,v}) \times P_{u,v}]$$

$$s' = s \oplus [F_{u,v} \times P_{u,v}] \oplus [(1-F_{u,v}) \times P_{u,v}]$$

Since $F_{u,v} \in \{0,1\}$, it follows from (3.4) that

$$s' = s \oplus P_{u,v} = s \oplus s \oplus b = b$$

Thus we obtain condition (3.1) and correctness of the data hiding scheme is proven.

*E. Algorithm 3*

To improve the safety level of the Algorithm 2, we can use an integer number $q \in \{0,1,2,\ldots,2^r - 1\}$ as a second key. We calculate Algorithm 3 with content similar to the Algorithm 2 except value s is calculated by the formula:

$$s = XSUM(F \otimes P) \oplus q$$

Additionally, to restore the bit string b, instead of the formula (3.1) we will use the following formula:

$$XSUM(G \otimes P) \oplus q = b$$

We notice that matrix G in Algorithm 3 is determined from F, P, q and b. Therefore, we can see that this algorithm as a transformation T from (F, P, q, b) to G:

$$G = T(F,P,q,b)$$

## IV. DATA HIDING SCHEME IN BINARY IMAGE

*A. The Inputs for scheme*

Below we present use of the Algorithm 3 to hide a data bit string d in a cover binary image I. To do this, we need to use a positive integer r, a matrix P of size m×n and a sequence Q of m×n integers, which satisfy the following conditions:

- $r \leq \lfloor \log_2(m \times n + 1) \rfloor$
- $P_{i,j} \in \{0,1,\ldots,2^r - 1\} \; \forall (i,j)$
- $\{1,2,\ldots,2^r-1\} \subset \{P_{i,j} \,|\, i=1,\ldots,m \text{ and } j=1,\ldots,n\}$
- $Q = \{q_1, q_2, \ldots, q_{m\times n}\}$ with $0 \leq q_i \leq 2^r - 1$

*B. Algorithm for hiding data*

**Step 1 (Partition):** Divide the binary image I into N blocks $F^i$ of size m×n and divide the data string d into N sub-strings $b^i$ of size r bits.

**Step 2 (Hiding data in each block):**

For i=1 to N do

$$\alpha = (i-1) \bmod (m \times n) + 1$$
$$G^i = T(F^i, P, q_\alpha, b^i)$$

End for

After executing the algorithm, we get the binary image J including N blocks $G^i$ of size m×n.

*C. Algorithm for restoring data*

To restore hidden data from the stego-image J (image contains hidden information) we need to know r, m, n and secret keys P, Q. The algorithm is implemented as follows:

**Step 1 (Partition):** Divide the stego-image J into N blocks $G^i$ of size m×n.

**Step 2 (Restoring data):**

For i = 1 to N do

$$\alpha = (i-1) \bmod (m \times n) + 1$$
$$b^i = XSUM(G^i \otimes P) \oplus q_\alpha$$

End for

After executing the algorithm, we obtain data string d including N sub-strings $b^i$ of size r bits.

*D. Security Analysis of the Proposed Scheme*

Each data hiding scheme often uses matrices and/or number sequences as a secret key to protect the hidden data. The greater the number of key combinations, the more difficult it is for hackers to detect the secret key used. Therefore the scheme is of higher security.

The TCP scheme uses a binary m×n matrix K and a weight m×n matrix W as the secret keys. The number of combinations for K is $2^{m\times n}$ and for W is:

$$C_{2^r-1}^{m\times n} \times (2^r - 1)! \times (2^r - 1)^{m\times n-(2^r-1)}$$

So the number of key combinations (K, W) is:

$$\pi_{KW} = 2^{m\times n} \times C_{2^r-1}^{m\times n} \times (2^r - 1)! \times (2^r - 1)^{m\times n-(2^r-1)}$$

In [1], the authors use a binary m×n matrix K and a serial number m×n matrix O as the secret keys. Moreover, the authors pointed out that the number of combinations for O is:

$$C_{2^r-1}^{m\times n} \times (2^r - 1)! \times (2^r)^{m\times n-(2^r-1)}$$

So the number of key combinations (K, O) is:

$$\pi_{KO} = 2^{m\times n} \times C_{2^r-1}^{m\times n} \times (2^r - 1)! \times (2^r)^{m\times n-(2^r-1)}$$

In the proposed scheme we use an integer m×n matrix P and a sequence Q of m×n integer numbers as the secret keys. From the definition of P and Q in subsection IV.A, it follows that the number of combinations for P is:

$$C_{2^r-1}^{m\times n} \times (2^r - 1)! \times (2^r)^{m\times n-(2^r-1)}$$

and for Q is $2^{r \times m \times n}$, so the number of key combinations (P, Q) is:

$$\pi_{PQ} = 2^{r \times m \times n} \times C_{2^r-1}^{m \times n} \times (2^r - 1)! \times (2^r)^{m \times n - (2^r - 1)}$$

In applications often choose r ≥ 2, so we have:

$$\frac{\pi_{PQ}}{\pi_{KW}} > 2^{(r-1) \times m \times n} \geq 2^{m \times n}$$

$$\frac{\pi_{PQ}}{\pi_{KO}} = 2^{(r-1) \times m \times n} \geq 2^{m \times n}$$

The above analysis shows that the new proposed scheme is more secure than both schemes TCP and CTL

## V. EXPERIMENTS

In these experiments we use three different images of the same size 256×256 as cover images (Figure 1), including English text image, Vietnamese text image and the "Lena" image, to hide the same message with 256 bytes length (Figure 2). The data hiding in each image were performed according to two plans of dividing blocks: (m,n,r) = (8,8,6) and (m,n,r)= (16,16,8).

Table 1 presents the PSNR values of all stego-images obtained by the new scheme, the CTL scheme and the TCP scheme, respectively. The results indicate that, PSNR values of the new scheme are always higher than those of TCP scheme and the same as those of CTL scheme.

Table 2 presents number of pixels modified in each image after performing data hiding by above schemes. The results indicate that these numbers of the new scheme are always smaller than those of TCP scheme and the same as those of CTL scheme.

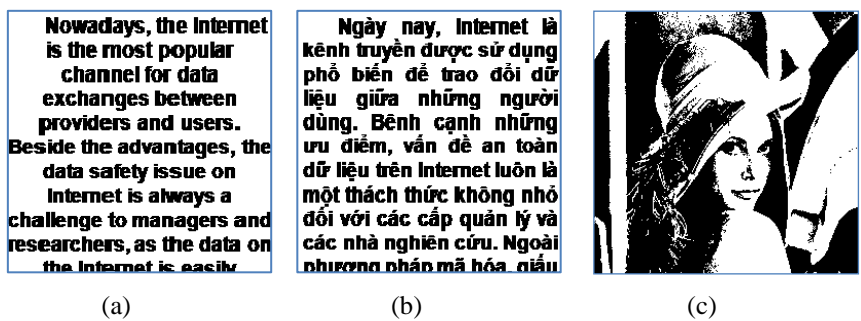

|     (a)     |     (b)     |     (c)     |

**Fig. 1.** Cover images: (a) English text image, (b) Vietnamese text image, (c) Lena image

> It is important to understand that cyber warfare does not necessarily have anything to do with the internet. Many of the more devastating cyber - attacks can not be launched remotely, as the most critical networks are not connected to the public network.

**Fig. 2.** The secret message with 256 characters

**Table 1.** PSNR values of stego-images of three schemes

| Block size / Cover Image | 8×8 | | | 16×16 | | |
|---|---|---|---|---|---|---|
|  | *New scheme* | *CTL scheme* | *TCP scheme* | *New scheme* | *CTL scheme* | *TCP scheme* |
| Vietnamese text image | 22,901 dB | **22,94 dB** | 21,83 dB | 24,116 dB | **24,134 dB** | 23,196 dB |
| English text image | **22,94 dB** | **22,94 dB** | 22,005 dB | **24,134 dB** | 24,116 dB | 23,1 dB |
| Lena image | **22,901 dB** | 22,889 dB | 22,166 dB | **24,151 dB** | **24,151 dB** | 22,967 dB |

**Table 2.** Number of modified pixels in stego images of three schemes

| Block size / Stego Images | 8×8 | | | 16×16 | | |
|---|---|---|---|---|---|---|
|  | *New scheme* | *CTL scheme* | *TCP scheme* | *New scheme* | *CTL scheme* | *TCP scheme* |
| Vietnamese text image | 336 bits | **333 bits** | 430 bits | 254 bits | **253 bits** | 314 bits |
| English text image | **333 bits** | **333 bits** | 413 bits | **253 bits** | 254 bits | 321 bits |
| Lena image | **336 bits** | 337 bits | 398 bits | **252 bits** | **252 bits** | 331 bits |

## VI. CONCLUSIONS

This paper presents a new scheme for embedding secret data into a binary image. For each block of m × n pixels, the new scheme can hide $\lfloor log_2(m \times n + 1) \rfloor$ bits of data by changing one bit at most in block. The experimental results indicate that if embedding a same amount of secret data in a same cover image, the stego-image quality of the new scheme is similar to that of CTL scheme and better than that of TCP scheme. The theoretical analyses have confirmed that the new proposed scheme is indeed more secure than both schemes TCP and CTL. Additionally, as compared to two schemes above, the new scheme is simpler and easier to install for applications.

## REFERNCES

## AUTHORS PROFILE

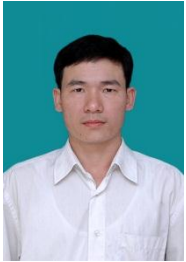

**Do Van Tuan** received M.Sc. degree in Information Technology in 2007 from Vietnam National University, Ha Noi. He is currently a PhD student at Hanoi University of Science and Technology. His research interests include Data Hiding, Digital Watermarking, Cryptography

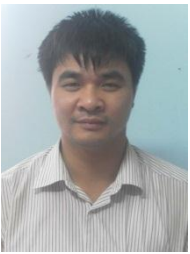

**Tran Dang Hien** received M.Sc. degree in Information Technology in 2010 from Vietnam National University, Ha Noi. He is currently a PhD student at Vietnam National University. His research interests include Data Hiding, Digital Watermarking, Image Forensic.

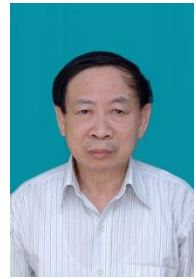

**Pham Van At** received B.Sc. and PhD degree in Mathematics in 1967 and 1980 from Vietnam National University, Ha Noi. Since 1984 he is Associate Professor at Faculty of Information Technology of Hanoi University of Transport and Communication. His research interests include Linear algebra, optimization, Image processing, Data Hiding, Cryptography.